\pdfoutput=1
\documentclass{jdfsl}
\usepackage{makeidx}  
\usepackage{upgreek}
\usepackage{url}
\usepackage{amsmath}
\usepackage[T1]{fontenc}
\usepackage{array}
\usepackage{subfig}
\usepackage{multirow}
\usepackage{array}
\usepackage{amssymb}
\usepackage{natbib}
\usepackage{enumitem}
\usepackage{amssymb}
\usepackage{pifont}
\usepackage{textcomp}

\usepackage{upgreek}
\usepackage{url}
\usepackage{amsmath}
\usepackage{graphicx}
\usepackage[T1]{fontenc}
\usepackage{array}
\usepackage{color,soul}

\graphicspath{{./pdf/}{./jpeg/}{./images/}}
   \DeclareGraphicsExtensions{.pdf,.jpeg,.jpg,.png}

\begin{document}

\title{Current Challenges and Future Research Areas for Digital Forensic Investigation}

\author{David Lillis, Brett A. Becker, Tadhg O'Sullivan and Mark Scanlon}

\institute{~ \\ School of Computer Science,\\ University College Dublin, Ireland.\\ \{david.lillis, brett.becker, t.osullivan, mark.scanlon\}@ucd.ie, }

\abstract{
Given the ever-increasing prevalence of technology in modern life, there is a corresponding increase in the likelihood of digital devices being pertinent to a criminal investigation or civil litigation. As a direct consequence, the number of investigations requiring digital forensic expertise is resulting in huge digital evidence backlogs being encountered by law enforcement agencies throughout the world. It can be anticipated that the number of cases requiring digital forensic analysis will greatly increase in the future. It is also likely that each case will require the analysis of an increasing number of devices including computers, smartphones, tablets, cloud-based services, Internet of Things devices, wearables, etc. The variety of new digital evidence sources poses new and challenging problems for the digital investigator from an identification, acquisition, storage and analysis perspective. This paper explores the current challenges contributing to the backlog in digital forensics from a technical standpoint and outlines a number of future research topics that could greatly contribute to a more efficient digital forensic process.
}


\maketitle

\section{Introduction}


The early 21st century has seen a dramatic increase in new and ever-evolving technologies available to consumers and industry alike. Generally, the consumer-level user base is now more adept and knowledgeable about what technologies they employ in their day-to-day lives. The number of cases where digital evidence is relevant to an investigation is ever increasing and it is envisioned that the existing backlog for law enforcement will balloon in the coming years as the prevalence of digital devices increases. It is for these reasons that it is important to take stock of the current state of affairs in the field of digital forensics. Cloud based services, Internet-of-Things devices, anti-forensic techniques, distributed and high capacity storage, and the sheer volume and heterogeneity of pertinent devices pose new and challenging problems for the acquisition, storage and analysis of this digital evidence. 

Due to the sheer volume of data to be acquired, stored, analysed and reported on, combined with the level of expertise necessary to ensure the court admissibility of the resultant evidence, it was inevitable that a significant backlog in cases awaiting analysis would occur \citep{hitchcock2016tiered}. Three particular aspects have contributed to this backlog~\citep{quick2014impacts}:

\begin{enumerate}
\item An increase in the number of devices that are seized for analysis per case.
\item The number of cases whereby digital evidence is deemed pertinent is ever increasing.
\item The volume of potentially evidence-rich data stored on each item seized is also increasing.
\end{enumerate}

This backlog is having a significant impact on the ideal legal process. According to a report by the \citet{gardai2015} (Irish National Police), delays of up to four years in conducting digital forensic investigations on seized devices have ``seriously impacted on the timeliness of criminal investigations'' in recent years. In some cases, these delays have resulted in prosecutions being dismissed in courts. This issue regarding the digital evidence backlog is further compounded due to the cross-border, intra-agency cooperation required by many forensic investigations. If a given country has an especially low digital investigative capacity, it can have a significant knock-on effect in an international context \citep{james2014measuring}.

In this paper, we review relevant recent research literature to elucidate the developments and current challenges in the field. While much progress has been made in the digital forensic process in recent years, little work has made appreciable progress in tackling the evidence backlog in practice. While evidence is lying unanalysed in an evidence store, investigations are often left waiting for new leads to be discovered, which has serious consequences for following these new threads of investigation at a later date. A number of practical infrastructural improvements to the current forensic process are discussed including automation of device acquisition and analysis, Forensics-as-a-Service (FaaS), hardware-facilitated heterogeneous evidence processing, remote evidence acquisition, and cross-jurisdictional evidence sharing over the Internet. These infrastructural improvements will enable a number of both new and improved forensic processes. These may include data visualisation, multi-device evidence and timeline resolution, data deduplication for storage and acquisition purposes, parallel or distributed investigations and process optimisation of existing techniques. The aforementioned improvements should combine to aid law enforcement and private digital investigators to greatly expedite the current forensic process. It is envisioned that the future research areas presented as part of this paper will influence further research in the field.


\section{Current Challenges}

\citet{raghavan2013digital} outlined five major challenge areas for digital forensics, gathered from a survey of research in the area:

\begin{enumerate}
	\item The complexity problem, arising from data being acquired at the lowest (i.e. binary) format with increasing volume and heterogeneity, which calls for sophisticated data reduction techniques prior to analysis.
	\item The diversity problem, resulting naturally from ever-increasing volumes of data, but also from a lack of standard techniques to examine and analyse the increasing numbers and types of sources, which bring a plurality of operating systems, file formats, etc. The lack of standardisation of digital evidence storage and the formatting of associated metadata also unnecessarily adds to the complexity of sharing digital evidence between national and international law enforcement agencies \citep{scanlon2014p2pevidencebag}.
	\item The consistency and correlation problem resulting from the fact that existing tools are designed to find fragments of evidence, but not to otherwise assist in investigations.
	\item The volume problem, resulting from increased storage capacities and the number of devices that store information, and a lack of sufficient automation for analysis.
	\item The unified time lining problem, where multiple sources present different time zone references, timestamp interpretations, clock skew/drift issues, and the syntax aspects involved in generating a unified timeline.
\end{enumerate}

Numerous other researchers have identified more specific challenges, which can generally be categorised according to Raghavan's above classification. Examples include \cite{Garfinkel2010}, \cite{Wazid2013}, and \cite{karie2015taxonomy}.

It is widely agreed that the volume of data that is potentially relevant to investigations is growing rapidly. The amount of data per case at the FBI's 15 regional computer forensic laboratories has grown 6.65 times between 2003-2011, from 84GB to 559GB \citep{roussev2013real}. One cause of this is the growth in storage capacities that has occurred in recent years. Additionally, the increasing proliferation of mobile and 
(IoT) devices adds to the number of devices that require examination in a given investigation. Beyond the magnitude of the data, the use of cloud services means that it may not be clear what data exists and where it is actually located. 

As advanced mobile and wearable technologies have continued to become more ubiquitous amongst the general population, they also now play a more prevalent role in digital forensic investigations. Over the past decade the capabilities of these smart devices have reached a point where they can function at a level near to that of the average household computer and are currently only limited by processing power and storage capacity. This contributes to the diversity problem, where a greater variety of devices become candidates for digital forensic investigation (e.g. \citet{Baggili2015} has reported on forensics on smart watches). Mobile and IoT devices make use of a variety of operating systems, file formats and communication standards, all of which add to the complexity of digital investigations. In addition, embedded storage may not be easily removable from devices, unlike for traditional desktop and server computers, and in some cases a devices will lack persistent storage entirely, necessitating expensive RAM forensics.

Investigating multiple devices also contributes to the consistency and correlation problem, where evidence gathered from distinct sources must be correlated for temporal and logical consistency. This is often performed manually: a significant drain on investigators' resources. The requirements for RAM forensics also becomes pertinent in cases of anti-forensics, where a digital criminal takes measures to avoid evidence being acquired, including the creation of malware that resides in RAM alone. The increasing sophistication of digital criminals' activities is also a substantial challenge.

Other issues include limitations on bandwidth for transferring data for investigation, the volatility of evidence, the fact that digital media has a limited lifespan that may possibly result in evidence being lost, and the increasing ubiquity of encryption in modern communications and data storage.

The following sections concentrate on a number of important emerging trends in modern computing that contribute to the problems outlined above.

\subsection{Internet-of-Things} \label{IoT}

The Internet-of-Things (IoT) refers to a vision of everyday items that are connected to a network and send data to one another. \cite{juniper2015IoT} estimate that there are already 13.4bn IoT devices in existence 2015, and they expect this figure to reach 38.5bn by 2020. These IoT devices are typically deployed in two broad areas: in the consumer domain (smart home, connected vehicles, digital healthcare) and in the industrial domain (retail, connected buildings, agriculture). Some IoT devices are commonplace items that have Internet connectivity added (e.g. refrigerators, TVs), whereas others are newer sensing or actuation devices that have been developed with the IoT specifically in mind. 

The IoT has the potential to become a rich source of evidence from the physical world, and as such it poses its own unique set of challenges for digital forensic investigators~\citep{Hegarty2014}. Compared to traditional digital forensics, there is less certainty in where data originated from, and where it is stored. Data persistence may be a problem. IoT devices themselves typically have limited memory (and may have no persistent data storage). Thus any data that is stored for longer periods may be stored in some in-network hub, or sent to the cloud for more persistent storage. This therefore means that the challenges related to cloud forensics (as discussed below in Section~\ref{cloud}) will likely apply in the IoT domain also.

Already, some efforts have begun to analyse IoT devices for forensics purposes (e.g. \cite{Sutherland2014} on smart TVs), however this work is in its early stages at present. The heterogeneous nature of IoT devices, including differences in operating systems, filesystems and communication standards, adds significantly to the complexity, diversity and correlation problems for forensic investigators.

\cite{Ukil2011} outline some security concerns of IoT researchers, which feed directly into the desires of forensic investigators, incorporating issues such as availability, authenticity and non-repudiation, which are important for legally-sound use of the data. These are addressed using encryption technologies, which are easy to incorporate into computationally powerful devices that are connected to mains energy. However it becomes more of a challenge for smaller, battery-operated, computationally-constrained devices, where such considerations may be sacrificed. This has inevitable consequences for the usefulness of the data in a legal context.

\subsection{Emerging Cloud Computing or Cloud Forensic Challenges} \label{cloud}

Usage of cloud services such as Amazon Cloud Drive, Office 365, Google Drive and Dropbox are now commonplace amongst the majority of Internet users.  From a digital forensics point of view, these services present a number of unique challenges, as has been reported in the 2014 National Institute of Standards and Technology's draft report \citep{NIST}. Typically, data in the cloud is distributed over a number of distinct nodes unlike more traditional forensic scenarios where data is stored on a single machine.  Due to the distributed nature of cloud services, data can potentially reside in multiple legal jurisdictions, leading to investigators relying on local laws and regulations regarding the collection of evidence~\citep{SimouCloud, Ruan201334}. This can potentially increase the time, cost and difficulty associated with a forensic investigation. From a technical standpoint, the fact that a single file can be split into a number of data blocks that are then stored on different remote nodes adds another layer of complexity thereby making traditional digital forensic tools redundant \citep{7069509, 6544395}.

Additionally, the Cloud Service Providers (CSP) and their user base must be taken into consideration.  Investigators are reliant on the willingness of CSPs to allow for the acquisition and reproduction of data. The lack of standardisation among the varying CSPs, differing levels of data security and their Service Level Agreements are obstacles to both cloud forensic researchers and investigators \citep{6544395}. The multi-tenancy of many cloud systems poses three significant challenges to digital forensic investigations. In the majority of cases the privacy and confidentiality of legitimate users must be taken into account by investigators due to the shared infrastructures that support cloud systems \citep{7113537}. The distributed nature of cloud systems along with multi-tenancy can require the acquisition of vast volumes of data leading to many of the challenges outlined below. Finally, the use of IP anonymity and the easy-to-use features of many cloud systems, such as requiring minimal information when signing up for a service, can lead to situations where identifying a criminal is near impossible \citep{6418812, Ruan201334}.

Cloud forensics also faces a number of challenges associated with traditional digital forensic investigations. Encryption and other anti-forensic techniques are commonly used in cloud-based crimes. The limited time for which forensically-important data is available is also an issue with cloud-based systems. Due to the fact that said systems are continuously running data, can be overwritten at any time.  Time of acquisition has also proved a challenging task in regard to cloud forensics. \citet{NehaCloud} showed that commonly-used forensic tools such as the Linux \textbf{\texttt{dd}} command and Amazon's AWS Snapshot took a considerable amount of time to acquire 30Gb of data from a cloud service.

While advances continue with regard to the tools and techniques used in cloud forensics, the aforementioned challenges continue to impede investigations. \citet{HenrySans} produced results showing that investigations on cloud-based systems make up only a fraction of all digital forensic investigations.  Many investigations are stalled beyond the point of a perpetrator's owned devices and rarely extend into the cloud-based services they use. Results such as these form a strong argument for continued research in this field.

\section{Future Research}

\subsection{Distributed Processing}

Distributed Digital Forensics has been discussed for some time \citep{roussev2004breaking,shanmugasundaram2003fornet,garfinkel2009bringing, beebe2009digital}. However there is more scope for it to be put into practice. \citet{roussev2013real} cite two main reasons that the processing speed of current generation digital forensic tools is inadequate for the average case: First, users have failed to formulate explicit performance requirements and second, developers have failed to put performance as a top-level concern in line with reliability and correctness. They proposed and validated a new approach to target acquisition that enables file-centric processing without disrupting optimal data throughput from the raw device. Their evaluation of core forensic processing functions with respect to processing rates shows intrinsic limitations in both desktop and server scenarios. Their results suggest that with current software, keeping up with a commodity SATA HDD at 120 MB/s requires between 120 and 200 cores. 

\subsection{HPC and Parallel Processing}
Despite the bottleneck of many digital forensic operations being disk read speed, there are steps in the process that are not limited by the physical read speed of the storage device. For instance the analysis phase can consume large amounts of time by computers and humans. High performance computing (HPC) advantages should be employed wherever possible to reduce computation time, and in an effort to reduce the time required by humans. Traditional HPC techniques normally exploit some level of parallelism, and to date have been underexploited by the digital forensic community. There are many applications where HPC techniques and hardware could be employed, for instance on expediting each part of the digital forensic process after the acquisition phase, i.e., preprocessing, storage, analysis and reporting.

\subsection{GPU-Powered Multi-threading}
GPUs excel at ``single instruction, multiple data'' (SIMD) computations with large numbers of general-purpose stream processors that can execute massively threaded algorithms for a number of applications and stand to do so for many digital forensics requirements in theory. 

\citet{marziale2007massive}, noted that GPUs have traditionally been both difficult to program and targeted at very specific problems. More recently, multicore CPUs coupled with GPU accelerators have been widely used in high performance computing due to better power efficiency and performance/price ratio \citep{zhong2012data}. In addition, there is now a multitude of integrated GPUs that are on the same silicon die as the CPU, bringing both easier programming models and greater efficiency. 

With new heterogeneous architectures and programming models such as these, powerful and efficient computer systems can be found in workstations with transparent access to CPU virtual addresses and very low overhead for computation offloading, and \citet{power2015toward} have shown such architectures to be advantageous in analytic processing. These seem very well suited for many digital forensics applications, particularly as technologies such as SSDs reduce the I/O bottleneck.

Nonetheless, the use of GPUs in digital forensics is largely absent from the literature and there are few standard digital forensic tools that utilise GPU acceleration. \citet{marziale2007massive} measured the effectiveness of offloading processing typical to digital forensics tools (such as file carving) to GPUs and found significant performance gains compared to simple threading techniques on multicore CPUs. Although the programming of the GPUs was more complex, the authors found that the effort was worth the performance gains. \citet{collange2009using} researched the feasibility of employing GPUs to accelerate the detection of sectors from contraband files using sector-level hashes. 

Their application was able to inspect several disk drives simultaneously and asynchronously from each other. In addition, disks from different computers can be inspected independently by the application. This approach indicated that the use of GPUs is viable. 

However, \citet{zha2011fast} employed multi-pattern search algorithms to reduce the time needed for file carving with Scalpel, showing that the limiting factor for performance is disk read time. The authors state there is no advantage to using GPUs, at least until mechanisms to read the disk faster are found. However, this conclusion assumes only one disk, and the traditional digital forensic model. In the new era of cloud forensics, SSDs, and other technological evolutions, this I/O bottleneck will be much less restrictive. 

\citet{iacob2015gpu} have employed GPUs in information retrieval cases where response time is of importance, similarly to DF. They demonstrate significant speed-up of two Bloom filter operations, which are used in approximate matching forensic applications \citep{breitinger2014automated}.

GPUs, like many new technologies, present new considerations for digital forensics. \citet{bress2013forensics} researched the use of GPUs to process confidential/sensitive information and found that data in GPU RAM is retrievable by unauthorised users by creating a dump of device memory. However this does not impede the use of GPUs for processing confidential information when the system itself is only accessible to authorised users.  

\subsection{DFaaS}

Digital Forensics as a Service (DFaaS) is a modern extension of the traditional digital forensic process. Since 2010, the Netherlands Forensic Institute (NFI) have implemented a DFaaS solution in order to combat the volume of backlogged cases \citep{van2014digital}. This DFaaS solution takes care of much of the storage, automation, investigator enquiry in the cases it manages. \citet{van2014digital} describe the advantages of the current system including efficient resource management, enabling detectives to directly query the data, improving the turn around time between forming a hypothesis in an investigation its confirmation based on the evidence, and facilitating easier collaboration between detectives working on the same case through annotation and shared knowledge.

While the aforementioned DFaaS system is a significant step in the right direction, many improvements to the current model could greatly expedite and improve upon the current process. This includes improving the functionality available to the case detectives, improving its current indexing capabilities and on-the-fly identification of incriminating evidence during the acquisition process \citep{van2014digital}.

Seeing as the DFaaS model is a cloud-based, remote access model, two significant disadvantages to the model are potential latency in using the online platform and being dependant on the upload bandwidth available during the physical storage acquisition phase of the investigation. A deduplicated evidence storage system, such as that described by \citet{watkins2009teleporter}, would facilitate the faster acquisition with each unique file across a number of investigations only needing to be stored, indexed, analysed and annotated once on the system. Eliminating non-pertinent, benign files during the acquisition phase of the investigation would greatly reduce the acquisition time (e.g., operating system, application, previously acquired non-incriminating files, etc.). This could greatly expedite pertinent information being available to the detectives working on the case as early as possible in the investigation. In order for any evidence to be court admissible, a forensically sound entire disk image would need to be reconstructible from the deduplicated data store, improving upon the system proposed by \citet{watkins2009teleporter}. Employing such a system would also facilitate a cloud-to-cloud based storage event monitoring of virtual systems as merely the changes of the virtual storage would need to be stored between each acquisition.

\subsection{Field-programmable Gate Arrays}
FPGAs are integrated circuits that can be configured after manufacture. FPGAs can implement any function that application-specific integrated circuits can, and offer several advantages over traditional CPUs. FPGAs can exploit inherent algorithmic parallelism (including low-level parallelism), and can often achieve results in fewer logic operations compared to traditional general purpose CPUs, resulting in faster processing times. FPGAs have recently found application in areas such as digital signal processing, imaging and video applications, and cryptography. Despite demonstrating desirable traits for digital forensics researchers, they have yet to be exploited for non-I/O-bound facets of digital forensics. Furthermore, as SSDs and other technologies ease the I/O bottleneck, FPGAs stand to be more broadly applicable in digital forensics.

\subsection{Applying Complementary Cutting Edge Research to Forensics}

Current investigation practice involves the analysis of data on standalone workstations. As such, the sophistication of the techniques that can be practically employed are limited. Much research has been conducted in a variety of areas that has theoretical relevance to digital forensics, but has been impractical to apply to date. A movement towards DFaaS and high-performance computing, as discussed above, offers advantages beyond merely expediting the techniques currently used in forensics investigations, which remain reliant on manual input. It also promises a situation where this complementary research may practically be brought to bear on digital forensic investigations.

One such research area is that of Information Retrieval (IR). Traditionally, IR is concerned with identifying documents within a corpus that help to satisfy a user's ``information need''. Traditionally, IR researchers have been faced with the trade-off between the competing goals of precision (retrieving only relevant documents) and recall (retrieving all the relevant documents), whereby improving on one of these metrics typically results in a reduction in the other. In IR for legal purposes, recall has long been acknowledged as being the more important metric, given that a single missing relevant document could have serious consequences for the prosecution of a criminal case, the enforcement of a contract, etc. However, focussing on recall frequently results in an investigator being required to manually sift through a large quantity of non-relevant documents. This is in contrast to web search, for example, where users typically do not require all of the relevant documents to be retrieved, of which there may possibly be millions. Instead, a web searcher wishes to avoid wasting time on non-relevant material.

IR for digital forensics is often seen as a typical example of legal information retrieval (e.g. by \citet{Beebe2007}). Although this is certainly true at the point a case is being built for court, it could be argued that the level of recall required at the triage stage can be sacrificed somewhat for greater precision, in order to allow investigators make speedy decisions about whether a given device should be investigated fully. Thus there is the potential for configurable IR systems to be utilised in forensics investigations, whose focus will change depending on the stage of the investigation.

The primary advantage of applying IR techniques to digital investigations is that once the initial preprocessing stage has been completed, searches can be conducted extremely quickly. \cite{furnas1987vocabulary} has shown that less than 20\% of searchers choose the same keywords for topics they are interested in. This suggests that many queries must be run to achieve full recall, and also suggests that standard IR techniques such as query expansion and synonym matching could also be applied to increase recall.

However, increasing recall typically reduces precision by also retrieving non-relevant documents as false positives. There are a number of ways in which this problem can be alleviated. The use of the aforementioned data deduplication techniques would eliminate standard system files from consideration (\cite{beebe2007new} note that the word ``kill'' appears as a command in many system files). Additionally, common visualisation approaches such as ranking \citep{beebe2014ranking} and clustering \citep{beebe2011clustering} are likely to help investigators in their manual search of retrieved documents.

Another consideration is that event timeline reconstruction is extremely important in a criminal investigation~\citep{chabot2014event}. When constructing a timeline from digital evidence, some temporal data is readily available (e.g. chat logs, file modification times, email timestamps, etc.), although it should be acknowledged that even this is not without its own challenges. Within the IR community, much research has been conducted into the extraction of temporal information from unstructured text~\citep{campos2014survey}. This can be used to dramatically reduce the manual load on investigators in this area.

\section{Conclusion} \label{conclusion}

In this paper a number of current challenges in the field of digital forensics are discussed. Each of these challenges in isolation can hamper the discovery of pertinent information for digital investigators and detectives involved in a multitude of different cases requiring digital forensic analysis. Combined, the negative effect of these challenges can be greatly amplified. These issues alongside limited expertise and huge workloads has resulted in the digital evidence backlog increasing to the order of years for many law enforcement agencies worldwide. The predicted ballooning of case volume in the near future will serve to further compound the backlog problem -- particularly as the volume of evidence from non-traditional sources, such as cloud-based and Internet-of-Things sources, is also likely to increase. 

In terms of research directions, practices already in place in many Computer Science sub-disciplines hold promise for addressing these challenges including those in distributed, parallel, GPU and FPGA processing, and information retrieval. More intelligent deduplicated evidence data storage and analysis techniques can help eliminate the duplicated processing and duplicated expert analysis of previously  content. These research directions can be applied to the traditional digital forensics process to help combat the aforementioned backlog through more efficient allocation of precious digital forensic expert time through the improvement and expedition of the process itself.

\newpage

\bibliographystyle{plainnat}
\begin{flushleft}
\bibliography{expedited}

\begin{thebibliography}{45}
\providecommand{\natexlab}[1]{#1}
\providecommand{\url}[1]{\texttt{#1}}
\expandafter\ifx\csname urlstyle\endcsname\relax
  \providecommand{\doi}[1]{doi: #1}\else
  \providecommand{\doi}{doi: \begingroup \urlstyle{rm}\Url}\fi

\bibitem[Almulla et~al.(2013)Almulla, Iraqi, and Jones]{6544395}
S.~Almulla, Y.~Iraqi, and A.~Jones.
\newblock {Cloud Forensics: A Research Perspective}.
\newblock In \emph{Innovations in Information Technology (IIT), 2013 9th
  International Conference on}, pages 66--71, March 2013.

\bibitem[Baggili et~al.(2015)Baggili, Oduro, Anthony, Breitinger, and
  McGee]{Baggili2015}
Ibrahim Baggili, Jeff Oduro, Kyle Anthony, Frank Breitinger, and Glenn McGee.
\newblock {Watch What You Wear: Preliminary Forensic Analysis of Smart
  Watches}.
\newblock In \emph{2015 10th International Conference on Availability,
  Reliability and Security}, pages 303--311. IEEE, Aug 2015.
\newblock ISBN 978-1-4673-6590-1.

\bibitem[Beebe(2009)]{beebe2009digital}
Nicole Beebe.
\newblock {Digital Forensic Research: The Good, the Bad and the Unaddressed}.
\newblock In \emph{Advances in Digital Forensics V}, pages 17--36. Springer,
  2009.

\bibitem[Beebe and Dietrich(2007)]{beebe2007new}
Nicole Beebe and Glenn Dietrich.
\newblock {A New Process Model for Text String Searching}.
\newblock In \emph{Advances in Digital Forensics III}, pages 179--191.
  Springer, 2007.

\bibitem[Beebe and Clark(2007)]{Beebe2007}
Nicole~Lang Beebe and Jan~Guynes Clark.
\newblock {Digital Forensic Text String Searching: Improving Information
  Retrieval Effectiveness by Thematically Clustering Search Results}.
\newblock \emph{Digital Investigation}, 4\penalty0 (S1):\penalty0 49--54, 2007.

\bibitem[Beebe and Liu(2014)]{beebe2014ranking}
Nicole~Lang Beebe and Lishu Liu.
\newblock {Ranking Algorithms for Digital Forensic String Search Hits}.
\newblock \emph{Digital Investigation}, 11\penalty0 (S2):\penalty0 314--322,
  2014.

\bibitem[Beebe et~al.(2011)Beebe, Clark, Dietrich, Ko, and
  Ko]{beebe2011clustering}
Nicole~Lang Beebe, Jan~Guynes Clark, Glenn~B. Dietrich, Myung~S. Ko, and Daijin
  Ko.
\newblock {Post-Retrieval Search Hit Clustering to Improve Information
  Retrieval Effectiveness: Two Digital Forensics Case Studies}.
\newblock \emph{Decision Support Systems}, 51\penalty0 (4):\penalty0 732--744,
  2011.

\bibitem[Breitinger and Roussev(2014)]{breitinger2014automated}
Frank Breitinger and Vassil Roussev.
\newblock {Automated Evaluation of Approximate Matching Algorithms on Real
  Data}.
\newblock \emph{Digital Investigation}, 11:\penalty0 S10--S17, 2014.

\bibitem[Bre{\ss} et~al.(2013)Bre{\ss}, Kiltz, and
  Sch{\"a}ler]{bress2013forensics}
Sebastian Bre{\ss}, Stefan Kiltz, and Martin Sch{\"a}ler.
\newblock {Forensics on GPU Coprocessing in Databases--Research Challenges,
  First Experiments, and Countermeasures.}
\newblock In \emph{BTW Workshops}, pages 115--129. Citeseer, 2013.

\bibitem[Campos et~al.(2014)Campos, Dias, Jorge, and Jatowt]{campos2014survey}
Ricardo Campos, Ga{\"e}l Dias, Al{\'\i}pio~M Jorge, and Adam Jatowt.
\newblock {Survey of Temporal Information Retrieval and Related Applications}.
\newblock \emph{ACM Computing Surveys (CSUR)}, 47\penalty0 (2):\penalty0 15,
  2014.

\bibitem[Chabot et~al.(2014)Chabot, Bertaux, Kechadi, and
  Nicolle]{chabot2014event}
Yoan Chabot, Aur{\'e}lie Bertaux, Tahar Kechadi, and Christophe Nicolle.
\newblock {Event Reconstruction: A State of the Art}.
\newblock \emph{Handbook of Research on Digital Crime, Cyberspace Security, and
  Information Assurance}, page~15, 2014.

\bibitem[Chen et~al.(2012)Chen, Du, Qin, and Du]{6418812}
Guangxuan Chen, Yanhui Du, Panke Qin, and Jin Du.
\newblock Suggestions to digital forensics in cloud computing era.
\newblock In \emph{Network Infrastructure and Digital Content (IC-NIDC), 2012
  3rd IEEE International Conference on}, pages 540--544, Sept 2012.

\bibitem[Chen et~al.(2015)Chen, Xu, Yuan, and Shashidhar]{7069509}
Lei Chen, Lanchuan Xu, Xiaohui Yuan, and N.~Shashidhar.
\newblock {Digital Forensics in Social Networks and the Cloud: Process,
  Approaches, Methods, Tools, and Challenges}.
\newblock In \emph{Computing, Networking and Communications (ICNC), 2015
  International Conference on}, pages 1132--1136, Feb 2015.

\bibitem[Collange et~al.(2009)Collange, Dandass, Daumas, and
  Defour]{collange2009using}
Sylvain Collange, Yoginder~S Dandass, Marc Daumas, and David Defour.
\newblock Using graphics processors for parallelizing hash-based data carving.
\newblock In \emph{System Sciences, 2009. HICSS'09. 42nd Hawaii International
  Conference on}, pages 1--10. IEEE, 2009.

\bibitem[Furnas et~al.(1987)Furnas, Landauer, Gomez, and
  Dumais]{furnas1987vocabulary}
George~W. Furnas, Thomas~K. Landauer, Louis~M. Gomez, and Susan~T. Dumais.
\newblock {The Vocabulary Problem in Human-System Communication}.
\newblock \emph{Communications of the ACM}, 30\penalty0 (11):\penalty0
  964--971, 1987.

\bibitem[{Garda S\'ioch\'ana Inspectorate}(2015)]{gardai2015}
{Garda S\'ioch\'ana Inspectorate}.
\newblock {Changing Policing in Ireland}, November 2015.

\bibitem[Garfinkel et~al.(2009)Garfinkel, Farrell, Roussev, and
  Dinolt]{garfinkel2009bringing}
Simson Garfinkel, Paul Farrell, Vassil Roussev, and George Dinolt.
\newblock {Bringing Science to Digital Forensics with Standardized Forensic
  Corpora}.
\newblock \emph{Digital Investigation}, 6:\penalty0 S2--S11, 2009.

\bibitem[Garfinkel(2010)]{Garfinkel2010}
Simson~L Garfinkel.
\newblock {Digital Forensics Research: The Next 10 Years}.
\newblock \emph{Digital Investigation}, 7:\penalty0 S64--S73, 2010.

\bibitem[Hegarty et~al.(2014)Hegarty, Lamb, and Attwood]{Hegarty2014}
Robert~C. Hegarty, David~J. Lamb, and Andrew Attwood.
\newblock {Interoperability Challenges in the Internet of Things}.
\newblock In Paul Dowland, Steven Furnell, and Bogdan Ghita, editors,
  \emph{Proceedings of the Tenth International Network Conference (INC 2014)},
  pages 163--172. Plymouth University, 2014.

\bibitem[Henry et~al.(2013)Henry, Williams, and Wright]{HenrySans}
Paul Henry, Jacob Williams, and Benjamin Wright.
\newblock {The SANS Survey of Digital Forensics and Incident Response}.
\newblock In \emph{Tech Rep}, July 2013.

\bibitem[Hitchcock et~al.(2016)Hitchcock, Le-Khac, and
  Scanlon]{hitchcock2016tiered}
Ben Hitchcock, Nhien-An Le-Khac, and Mark Scanlon.
\newblock {Tiered Forensic Methodology Model for Digital Field Triage by
  Non-Digital Evidence Specialists}.
\newblock \emph{Digital Investigation}, 13\penalty0 (S1), 03 2016.
\newblock Proceedings of the Third Annual DFRWS Europe.

\bibitem[Iacob et~al.(2015)Iacob, Itu, Sasu, Moldoveanu, and
  Suciu]{iacob2015gpu}
Alexandru Iacob, Lucian Itu, Lucian Sasu, Florin Moldoveanu, and Constantin
  Suciu.
\newblock Gpu accelerated information retrieval using bloom filters.
\newblock In \emph{System Theory, Control and Computing (ICSTCC), 2015 19th
  International Conference on}, pages 872--876. IEEE, 2015.

\bibitem[James and Jang(2014)]{james2014measuring}
Joshua~I James and Yunsik~Jake Jang.
\newblock {Measuring Digital Crime Investigation Capacity to Guide
  International Crime Prevention Strategies}.
\newblock In \emph{Future Information Technology}, pages 361--366. Springer,
  2014.

\bibitem[{Juniper Research}(2015)]{juniper2015IoT}
{Juniper Research}.
\newblock {The Internet of Things: Consumer, Industrial \& Public Services
  2015-2020}, July 2015.

\bibitem[Karie and Venter(2015)]{karie2015taxonomy}
Nickson~M Karie and Hein~S Venter.
\newblock {Taxonomy of Challenges for Digital Forensics}.
\newblock \emph{Journal of Forensic Sciences}, 60\penalty0 (4):\penalty0
  885--893, 2015.

\bibitem[Marziale et~al.(2007)Marziale, Richard, and
  Roussev]{marziale2007massive}
Lodovico Marziale, Golden~G Richard, and Vassil Roussev.
\newblock {Massive Threading: Using GPUs to Increase the Performance of Digital
  Forensics Tools}.
\newblock \emph{Digital Investigation}, 4:\penalty0 73--81, 2007.

\bibitem[Morioka and Sharbaf(2015)]{7113537}
E.~Morioka and M.S. Sharbaf.
\newblock {Cloud Computing: Digital Forensic Solutions}.
\newblock In \emph{Information Technology - New Generations (ITNG), 2015 12th
  International Conference on}, pages 589--594, April 2015.

\bibitem[NIST(2014)]{NIST}
NIST.
\newblock {NIST Cloud Computing Forensic Science Challenges}.
\newblock 2014.

\bibitem[Power et~al.(2015)Power, Li, Hill, Patel, and Wood]{power2015toward}
Jason Power, Yinan Li, Mark~D Hill, Jignesh~M Patel, and David~A Wood.
\newblock {Toward GPUs Being Mainstream in Analytic Processing}.
\newblock 2015.

\bibitem[Quick and Choo(2014)]{quick2014impacts}
Darren Quick and Kim-Kwang~Raymond Choo.
\newblock {Impacts of Increasing Volume of Digital Forensic Data: A Survey and
  Future Research Challenges}.
\newblock \emph{Digital Investigation}, 11\penalty0 (4):\penalty0 273--294,
  2014.

\bibitem[Raghavan(2013)]{raghavan2013digital}
Sriram Raghavan.
\newblock {Digital Forensic Research: Current State of the Art}.
\newblock \emph{CSI Transactions on ICT}, 1\penalty0 (1):\penalty0 91--114,
  2013.

\bibitem[Roussev and Richard~III(2004)]{roussev2004breaking}
Vassil Roussev and Golden~G Richard~III.
\newblock {Breaking the Performance Wall: The Case for Distributed Digital
  Forensics}.
\newblock In \emph{Proceedings of the 2004 Digital Forensics Research Workshop
  (DFRWS)}, volume~94, 2004.

\bibitem[Roussev et~al.(2013)Roussev, Quates, and Martell]{roussev2013real}
Vassil Roussev, Candice Quates, and Robert Martell.
\newblock {Real-time Digital Forensics and Triage}.
\newblock \emph{Digital Investigation}, 10\penalty0 (2):\penalty0 158--167,
  2013.

\bibitem[Ruan et~al.(2013)Ruan, Carthy, Kechadi, and Baggili]{Ruan201334}
Keyun Ruan, Joe Carthy, Tahar Kechadi, and Ibrahim Baggili.
\newblock {Cloud Forensics Definitions and Critical Criteria for Cloud Forensic
  Capability: An Overview of Survey Results}.
\newblock \emph{Digital Investigation}, 10\penalty0 (1):\penalty0 34 -- 43,
  2013.

\bibitem[Scanlon and Kechadi(2014)]{scanlon2014p2pevidencebag}
Mark Scanlon and M-Tahar Kechadi.
\newblock {Digital Evidence Bag Selection for P2P Network Investigation}.
\newblock In \emph{Proceedings of the 7th International Symposium on Digital
  Forensics and Information Security (DFIS-2013)}, pages 307--314. Springer,
  Gwangju, South Korea, 2014.

\bibitem[Shanmugasundaram et~al.(2003)Shanmugasundaram, Memon, Savant, and
  Bronnimann]{shanmugasundaram2003fornet}
Kulesh Shanmugasundaram, Nasir Memon, Anubhav Savant, and Herve Bronnimann.
\newblock {ForNet: A Distributed Forensics Network}.
\newblock In \emph{Computer Network Security}, pages 1--16. Springer, 2003.

\bibitem[Simou et~al.(2014)Simou, Kalloniatis, Kavakli, and
  Gritzalis]{SimouCloud}
Stavros Simou, Christos Kalloniatis, Evangelia Kavakli, and Stefanos Gritzalis.
\newblock {Cloud Forensics Solutions: A Review}.
\newblock In Lazaros Iliadis, Michael Papazoglou, and Klaus Pohl, editors,
  \emph{Advanced Information Systems Engineering Workshops}, volume 178 of
  \emph{Lecture Notes in Business Information Processing}, pages 299--309.
  Springer International Publishing, 2014.
\newblock ISBN 978-3-319-07868-7.

\bibitem[Sutherland et~al.(2014)Sutherland, Read, and Xynos]{Sutherland2014}
Iain Sutherland, Huw Read, and Konstantinos Xynos.
\newblock {Forensic Analysis of Smart TV: A Current Issue and Call to Arms}.
\newblock \emph{Digital Investigation}, 11\penalty0 (3):\penalty0 175--178, sep
  2014.

\bibitem[Thethi and Keane(2012)]{NehaCloud}
Neha Thethi and Anthony Keane.
\newblock {Digital Forensics Investigations in the Cloud}.
\newblock In \emph{IEEE International Advance Computing Conference (IACC)},
  Sept 2012.

\bibitem[Ukil et~al.(2011)Ukil, Sen, and Koilakonda]{Ukil2011}
Arijit Ukil, Jaydip Sen, and Sripad Koilakonda.
\newblock {Embedded Security for Internet of Things}.
\newblock In \emph{2011 2nd National Conference on Emerging Trends and
  Applications in Computer Science}, pages 1--6. IEEE, mar 2011.
\newblock ISBN 978-1-4244-9578-8.

\bibitem[van Baar et~al.(2014)van Baar, van Beek, and van Eijk]{van2014digital}
RB~van Baar, HMA van Beek, and EJ~van Eijk.
\newblock {Digital Forensics as a Service: A Game Changer}.
\newblock \emph{Digital Investigation}, 11:\penalty0 S54--S62, 2014.

\bibitem[Watkins et~al.(2009)Watkins, McWhorte, Long, and
  Hill]{watkins2009teleporter}
Kathryn Watkins, Mike McWhorte, Jeff Long, and Bill Hill.
\newblock {Teleporter: An Analytically and Forensically Sound Duplicate
  Transfer System}.
\newblock \emph{Digital Investigation}, 6:\penalty0 S43--S47, 2009.

\bibitem[Wazid et~al.(2013)Wazid, Katal, Goudar, and Rao]{Wazid2013}
Mohammad Wazid, Avita Katal, RH~Goudar, and Smitha Rao.
\newblock {Hacktivism Trends, Digital Forensic Tools and Challenges: A Survey}.
\newblock In \emph{Information \& Communication Technologies (ICT), 2013 IEEE
  Conference on}, pages 138--144. IEEE, 2013.

\bibitem[Zha and Sahni(2011)]{zha2011fast}
Xinyan Zha and Sartaj Sahni.
\newblock {Fast in-Place File Carving for Digital Forensics}.
\newblock In \emph{Forensics in Telecommunications, Information, and
  Multimedia}, pages 141--158. Springer, 2011.

\bibitem[Zhong et~al.(2012)Zhong, Rychkov, and Lastovetsky]{zhong2012data}
Ziming Zhong, Vladimir Rychkov, and Alexey Lastovetsky.
\newblock {Data Partitioning on Heterogeneous Multicore and Multi-GPU Systems
  Using Functional Performance Models of Data-Parallel Applications}.
\newblock In \emph{Cluster Computing (CLUSTER), 2012 IEEE International
  Conference on}, pages 191--199. IEEE, 2012.

\end{thebibliography}
\end{flushleft}

\end{document}